\documentstyle[prl,preprint,aps]{revtex}
\begin{document}

\tighten

\title{Self Adjoint Extensions of Phase and Time Operators}
\author{G.~Gour\thanks{E-mail:~gilgour@phys.ualberta.ca},
F. C.~Khanna\thanks{E-mail:~khanna@phys.ualberta.ca},
M.~Revzen\thanks{E-mail:~revzen@physics.technion.ac.il}}
\address{Theoretical Physics Institute,
Department of Physics, University of Alberta,\\
Edmonton, Canada T6G 2J1\\
and\\
Physics Department, Technion-Israel Institute of Technology,\\ Haifa
32000, Israel}

\maketitle

\begin{abstract}
It is shown that any real and even function of the phase (time) operator 
has a self-adjoint extension and its relation to the general phase operator 
problem is analyzed.  
\end{abstract} 


Problems in the definition of the quantum phase were first addressed by Fritz 
London~\cite{London} in 1926. One year latter Dirac~\cite{Dirac} 
introduced an operator 
solution which was proved to be incomplete by Susskind and Glogower~\cite{Susskind}
(for history and measurements see~Nieto~\cite{Nieto}).
Since then a series of workers have made many attempts
to resolve the problem (for reviews see~\cite{Rev}).

In the present paper, the quantum phase and time problems are considered in the context of 
projective measurements (i.e. not in the formalism of the so-called POM or POVM 
observables~\cite{Shapiro}).
It is shown that {\it any} real and even function of the 
phase (time) operator, has a self adjoint extension. There are two subspaces in which these 
operators are self adjoint: the subspace consisting of all the even square-integrable 
functions of the phase and the subspace consisting all the odd ones. 
That is, the non-existence of self adjoint phase and time operators does not, necessarily, 
preclude self adjoint extensions of their absolute value.       

The quantum phase problem may be traced to two sources:\\ 
({\it i}) The spectrum of the phase is restricted to a finite interval which is chosen
in this paper, somewhat arbitrarily, to be $[-\pi ,\pi]$.\\
({\it ii}) The number operator (or equivalently the Hamiltonian of a simple harmonic 
oscillator) is bounded from below. 

It can be seen very easily why condition ({\it i}) leads to a problem.
The matrix elements of $[{\bf N},{\bf\Phi}]$ 
(${\bf N}$ and ${\bf\Phi}$ are the number and phase operators respectively) 
in the number state basis $|n\rangle$,
\begin{equation}
\langle n|[{\bf N},{\bf \Phi}]|n'\rangle=(n-n')\langle n|{\bf \phi}|n'\rangle
\label{npn}
\end{equation}
vanish for $n=n'$ because $|\langle n|{\bf\phi}|n\rangle|\leq\pi$. That is,
$[{\bf N},{\bf \Phi}]\neq i$.
However, as we will see in the following, this problem can be resolved. Therefore, we emphasize 
that it is condition ({\it ii}) which makes it impossible to define self adjoint phase or 
time operators. 

Consider the Hilbert space, ${\cal H}$, consisting of 
square integrable functions on the segment $-\pi\leq\theta\leq \pi$,
with scalar product given by
\begin{equation}
\langle u,v\rangle =\int_{-\pi}^{\pi}\bar{u}(\theta)v(\theta)d\theta\;.
\end{equation}
The operator ${\bf J}_{z}\equiv -i\frac{d}{d\theta}$ is defined over all the 
differentiable functions $v(\theta)$; it represents the angular momentum 
observable of a {\it plane} rotator. Since the spectrum of $J_{z}$ is 
{\it not} bounded from below, the analog to condition ({\it ii}) in this case
is not satisfied.

The operator ${\bf J}_{z}$ is not a self adjoint operator 
(see for example p.87 in~\cite{Peres}).
Denoting the adjoint of ${\bf J}_{z}$ by ${\bf J}_{z}^{*}$ and the complex
conjugate of $u(\theta)$ by $\bar{u}(\theta)$, we have 
\begin{equation}
\langle u,{\bf J}_{z}v\rangle -\langle{\bf J}_{z}^{*} u,v\rangle
=-i[\bar{u}(\pi)v(\pi)-\bar{u}(-\pi)v(-\pi)].
\end{equation}
Since $v(\pi)$ and $v(-\pi)$ are arbitrary, the above expression will vanish 
if, and only if, $\bar{u}(\pi)=\bar{u}(-\pi)=0$. That is, the domain of 
${\bf J}_{z}^{*}$ is smaller than that of ${\bf J}_{z}$.

The family of operators, ${\bf J}_{z}^{\alpha}\equiv -i\frac{d}{d\theta}$ with 
$0\leq\alpha\leq 1$, whose domains of definition, ${\cal H}_{\alpha}$,
\begin{equation}
u(\theta)\in {\cal H}_{\alpha}\;\;\iff\;\;u(\pi)=e^{2\pi i\alpha}u(-\pi),
\end{equation}
are all self-adjoint operators ({\it cf.} p.88 in~\cite{Peres}); they are 
the self-adjoint extensions of ${\bf J}_{z}$.

The {\it angle} operator,
\begin{equation}
{\bf\Theta}:\;\;\;{\bf\Theta}u(\theta)=\theta u(\theta),
\end{equation}
does not ``belong'' to ${\cal H}_{\alpha}$. This is the problem
that stems from condition~({\it i}) above. It is the fact that 
$-\pi\leq\theta\leq\pi$ which makes it impossible to find a common 
domain for ${\bf J}_{z}^{\alpha}$ and ${\bf\Theta}$.

In order to resolve this problem let us define a series of
differentiable functions, 
$f_{\varepsilon}(\theta)\in {\cal H}_{\alpha =0}$, such that 
$f_{\varepsilon}(\theta)=\theta$ for 
$-\pi\leq\theta < \pi -\varepsilon$, and for 
$\theta\geq \pi -\varepsilon$ the function smoothly goes to 
the value of $f_{\varepsilon}(-\pi)$, in the limit 
$\theta\rightarrow\pi$. Therefore, 
$f_{\varepsilon}(-\pi)=f_{\varepsilon}(\pi)$ with 
$\lim_{\varepsilon\rightarrow 0}f_{\varepsilon}(\theta)\rightarrow\theta$.

Now, the family of operators 
${\bf\Theta}_{\varepsilon}\equiv f_{\varepsilon}({\bf\Theta})$ are all ``belong''
to ${\cal H}_{\alpha}$; they are a good approximation of the angle operator,
${\bf\Theta}$, in the limit of small $\varepsilon$. 

The canonical commutation relations
\begin{equation}
[{\bf J}_{z}^{\alpha},{\bf\Theta}_{\varepsilon}]
=-if_{\varepsilon}^{\prime}({\bf\Theta}),
\end{equation} 
can not be taken at $\varepsilon =0$ due to the discontinuity of the function
$f_{\varepsilon =0}(\theta)$ at the point $\theta=\pi$. Nevertheless, sometimes it is 
convenient (but not rigorous) to write the canonical commutation relation of the plane 
rotator observables in the following form:
\begin{equation}
[{\bf J}_{z}^{\alpha},{\bf\Theta}_{\varepsilon =0}]
=-i\left(1-2\pi\delta({\bf\Theta}-\pi)\right),
\end{equation}
where ${\bf\Theta}_{\varepsilon =0}\neq {\bf\Theta}$ only at the point 
$\theta=\pi$. Denoting the eigenstates of ${\bf J}_{z}^{\alpha}$ by 
$|m \rangle$ and taking the matrix elements of both sides of the equation above
give the identity
\begin{equation}
(m-m')\langle m|{\bf\Theta}_{\varepsilon =0}|m' \rangle 
=-i\left(\delta _{m,m'}-\exp[i(m'-m)\pi]\right).
\end{equation}
Note that for $m=m'$ both sides of the equation are zero 
({\it cf}. Eq.~(\ref{npn})). In this way, the problems that follow from 
condition~({\it i}) are (partly) resolved; although the angle operator does not 
belong to ${\cal H}_{\alpha}$, any periodic function of it is a self-adjoint 
operator in ${\cal H}_{\alpha}$.
Therefore, we are left with the
problems that arise from condition~({\it ii}).   

Consider the Hilbert space, ${\cal H}_{r}$, consisting
of square integrable functions, $u(r)$, with $0\leq r <\infty$, 
$u(r=\infty)=0$ and inner product
\begin{equation}
\langle u,v\rangle =\int_{0}^{\infty}\bar{u}(r)v(r)dr 
\end{equation}
The radial momentum ${\bf p}_{r}=-id/dr$, which is defined over all the 
differentiable functions $v(r)$, is not a self-adjoint operator. It satisfies:
\begin{equation}
\langle u,{\bf p}_{r}v\rangle -\langle{\bf p}^{*}_{r} u,v\rangle
=i\bar{u}(0)v(0).
\end{equation}
Hence, if $v(0)$ is arbitrary, the domain of ${\bf p}^{*}_{r}$ must be restricted
by the boundary condition $u(0)=0$. But, if $u(r)$ belongs to the domain of 
${\bf p}^{*}_{r}$ (i.e. $u(0)=0$) then also ${\bf p}^{*}_{r}u(r)=-iu^{\prime}(r)$ belongs
to the domain of ${\bf p}^{*}_{r}$. That is, also $u^{\prime}(r=0)=0$, and, by induction, the
$n$th derivative of $u(r)$ must be zero at $r=0$ ($u^{(n)}(r=0)=0$). 

Thus, it is possible to define a {\it self-adjoint} radial momentum (or time) operator only on 
the subspace ${\cal A}$ of the Hilbert space ${\cal H}_{r}$, where $u(r)\in {\cal A}$ if, and 
only if, all the derivatives of $u(r)$ vanish at $r=0$. For example, the square integrable function 
$u(r)=\exp\left(-{1\over r}\right)/r^{2}$ belongs to ${\cal A}$.

From a physical point of view, the domain ${\cal A}$ of the {\it self-adjoint} time operator, 
${\bf T}\equiv -id/dr$, is too small. Most of the physical wave functions do not belong to 
${\cal A}$. Therefore, we would like to augment the domain of the radial momentum (time) operator.
The price that we have to pay is that we have to work now only with even functions of ${\bf p}_{r}$.

Let us define two subspaces ${\cal H}_{r}^{+}$ and ${\cal H}_{r}^{-}$ of the Hilbert space
${\cal H}_{r}$ as follows:
\begin{equation}
u(r)\in {\cal H}_{r}^{+}\;\iff\;u^{(2m+1)}(r=0)=0\;\;{\rm for}\;{\rm all}\;\;m=0,1,2,...
\end{equation} 
and, similarly,
\begin{equation}
u(r)\in {\cal H}_{r}^{-}\;\iff\;u^{(2m)}(r=0)=0\;\;{\rm for}\;{\rm all}\;\;m=0,1,2,...
\end{equation} 
Note that all the even functions belong to ${\cal H}_{r}^{+}$ and all the odd functions
belong to ${\cal H}_{r}^{-}$. However, there are also non-even functions that belong to 
${\cal H}_{r}^{+}$ and non-odd functions that belong to ${\cal H}_{r}^{-}$. For example, 
${\cal A}={\cal H}_{r}^{+}\cap {\cal H}_{r}^{-}$.

{\it Theorem}: Let $f(r)$ be a function that can be expressed as a Taylor series around
the point $r=0$. Then, the operator ${\bf F}\equiv f({\bf p}_{r})$ has two self-adjoint 
extensions, with domains ${\cal H}_{r}^{+}$ and ${\cal H}_{r}^{-}$, if, and only if, 
$f(r)$ is a real and even function.

{\it Proof}: First, let us take ${\bf F}= {\bf p}_{r}^{2}=-d^{2}/dr^{2}$. Then we have,
\begin{equation}
\langle u,{\bf F}v\rangle -\langle {\bf F}^{*} u,v\rangle
=\bar{u}(0)\frac{dv}{dr}(0)-\frac{d\bar{u}}{dr}(0)v(0).
\end{equation} 
Hence, if we want ${\bf F}$ to be a self-adjoint operator, we must require 
$\bar{u}(0)=v(0)=0$ or $\frac{dv}{dr}(0)=\frac{d\bar{u}}{dr}(0)=0$. But, if, 
the domain of ${\bf F}$ and ${\bf F}^{*}$ is restricted by the boundary 
condition $\bar{u}(0)=v(0)=0$ (or $\frac{dv}{dr}(0)=\frac{d\bar{u}}{dr}(0)=0$),
then also the functions ${\bf F}u(r)=-u^{\prime\prime}(r)$ must satisfy the same 
boundary condition , i.e. $u^{\prime\prime}(r=0)=0$ (or $u^{\prime\prime\prime}(r=0)=0$).
Therefore, ${\bf p}_{r}^{2}$ is a self-adjoint operator if its domain is ${\cal H}_{r}^{+}$ 
or ${\cal H}_{r}^{-}$. It is easy to see that any real function of ${\bf p}_{r}^{2}$ also has 
the same self-adjoint extensions and any odd function of ${\bf p}_{r}$ does {\it not} have a 
self-adjoint extension $\Box$

We are now equipped with the tools to define self-adjoint operators that represent
even and real functions of the quantum phase. In order to do so we have to include 
condition~({\it i}) to the analysis above. This can be done by replacing the continuous  
variable $r$ with the integer variable $n$ ($n=0,1,2,...$). After this
transformation ${\bf p}_{r}^{2}\rightarrow {\bf\Phi}^{2}$, where ${\bf\Phi}$ denotes the 
phase operator.

Any even function $u(r)\in {\cal H}_{r}^{+}$ and odd function
$v(r)\in {\cal H}_{r}^{-}$ can be written as
\begin{equation}
u(r)=\int_{-\infty}^{\infty}d\chi U_{\chi}\cos(\chi r)\;\;{\rm and}\;\;
v(r)=\int_{-\infty}^{\infty}d\chi V_{\chi}\sin(\chi r),    
\end{equation}
where $U_{\chi}$ and $V_{\chi}$ are the Fourier components. After the
transformation $r\rightarrow n$, the analogs of $u(r)$ and $v(r)$ are given by
\begin{equation}
u(n)=\int_{-\pi}^{\pi}d\phi U_{\phi}\cos(\phi n)\;\;{\rm and}\;\;
v(n)=\int_{-\pi}^{\pi}d\phi V_{\phi}\sin(\phi n)
\label{aaa}
\end{equation}
where $u(n)\in {\cal H}_{N}^{+}$ and $v(n)\in {\cal H}_{N}^{-}$. The subspaces
${\cal H}_{N}^{+}$ and ${\cal H}_{N}^{-}$ are the analogs of ${\cal H}_{r}^{+}$ 
and ${\cal H}_{r}^{-}$. Note that the inner product of $u(n)$ and $v(n)$ is given 
by
\begin{equation}
\langle u|v\rangle=\sum_{n=0}^{\infty}\bar{u}(n)v(n).
\end{equation}

Consider the subspace ${\cal H}_{N}^{+}$. Since all the vectors $u(n)$ in 
${\cal H}_{N}^{+}$ can be written as in Eq.~(\ref{aaa}), the basis
\begin{equation}
e_{\phi}^{+}(n>0)\equiv \sqrt{\frac{2}{\pi}}\cos(\phi n)\;,\;\;
e_{\phi}^{+}(n=0)\equiv \frac{1}{\sqrt{\pi}}
\end{equation}
spans ${\cal H}_{N}^{+}$. It is normalized as follows:
\begin{equation}
\int_{0}^{\pi}d\phi\;e_{\phi}^{+}(n)e_{\phi}^{+}(n')=\delta _{n,n'}\;\;{\rm and}\;\;
\sum_{n=0}^{\infty}e_{\phi}^{+}(n)e_{\phi '}^{+}(n)=\delta(\phi-\phi ').
\end{equation}
Thus, in Dirac notation we have
\begin{equation}
|\phi\rangle =\frac{1}{\sqrt{\pi}}|n=0\rangle+\sqrt{\frac{2}{\pi}}
\sum_{n=1}^{\infty}\cos(n\phi)|n\rangle
\label{phi}
\end{equation}
where $|\phi\rangle$ is an eigenstate of any real and even function of the 
phase operator. It is the analog of $1/\sqrt{2}(|\theta\rangle+|-\theta\rangle)$.

In ${\cal H}_{N}^{+}$ any real and even function of the phase ${\bf\Phi}$ is
a self-adjoint operator. In particular, the absolute value of the quantum phase 
operator is given by
\begin{equation}
\left|{\bf\Phi}\right|= \int_{0}^{\pi}\phi\;|\phi\rangle\langle\phi|\;d\phi,
\label{p1}
\end{equation}
where $|\phi\rangle$ has been defined in Eq.~(\ref{phi}).
It is interesting to calculate the expectation values of $\left|{\bf\Phi}\right|$ 
in a coherent state 
\begin{equation}
|\gamma\rangle\equiv\exp(-\frac{1}{2}|\gamma|^{2})
\sum_{n=0}^{\infty}\frac{\gamma^{n}}{\sqrt{n!}}|n\rangle, 
\end{equation}
where $\gamma=\sqrt{N}e^{i\theta}$ is the eigenvalue of the annihilation
operator ${\bf a}$. 
It can be shown~\cite{Gour} that in the classical limit 
$N=|\gamma|^{2}\rightarrow\infty$,
\begin{equation}
\langle\gamma|{\bf\Phi}|\gamma\rangle \rightarrow  \frac{\pi}{2}
-\frac{4}{\pi}\sum_{s=1,3,5...}\frac{\cos(\theta s)}{s^{2}} 
=|\theta|\;.
\end{equation}
This result proves useful for establishing that $\left|{\bf\Phi}\right|$ has 
the correct large-field correspondence limit.

Carruthers and Nieto~\cite{CandN}, have defined the ``phase cosine'' 
${\bf C}\equiv\frac{1}{2}({\bf E}+{\bf E}^{\dag})$ and ``phase sine'' 
${\bf S}\equiv\frac{1}{2i}({\bf E}-{\bf E}^{\dag})$ operators, where
\begin{equation}
{\bf E}\equiv\sum_{n=0}^{\infty}|n\rangle\langle n+1|\;,
\end{equation}
represents the phase exponent. It can be seen very easily that ${\bf C}$ and ${\bf S}$ 
do not commute. Therefore, ${\bf C}$ and ${\bf S}$ cannot represent the sine and cosine 
of the phase. As we have shown here, only the cosine of the phase is a self-adjoint operator 
in ${\cal H}_{N}^{+}$; it is given by
\begin{equation}  
\cos{\bf\Phi} \equiv \int_{0}^{\pi}\cos\phi\;|\phi\rangle\langle\phi|\;d\phi
={\bf C}+\frac{1}{2}(\sqrt{2}-1)(|0\rangle\langle 1|+|1\rangle\langle 0|),
\label{cands}
\end{equation}
where the projectors involving number eigenstates 
$|0\rangle$ and $|1\rangle$ can be neglected 
for states with $\langle {\bf N}\rangle\gg 1$.

The subspace ${\cal H}_{N}^{-}$ is spanned by the basis
\begin{equation}
e_{\phi}^{+}(n)\equiv \sqrt{\frac{2}{\pi}}\sin(\phi n).
\end{equation}
We have shown that any even and real function of the phase is 
a self-adjoint operator in ${\cal H}_{N}^{-}$. In Dirac notation,
\begin{equation}
|\phi\rangle =\sqrt{\frac{2}{\pi}}\sum_{n=1}^{\infty}\sin(n\phi)|n\rangle,
\end{equation}
where $|\phi\rangle$ is an eigenstate of the {\it absolute} value of the 
phase operator. It is the analog of 
$1/\sqrt{2}(|\theta\rangle-|-\theta\rangle)$.
The quantum absolute phase observable is give by
\begin{equation}
\left|{\bf\Phi}\right|= \int_{0}^{\pi}\phi\;|\phi\rangle\langle\phi|\;d\phi\;,
\label{p2}
\end{equation}
and the cosine of the phase is given by
\begin{equation}
\cos\left|{\bf\Phi}\right|={\bf C}-(|0\rangle\langle 1|+|1\rangle\langle 0|)/2.
\end{equation}

In order to understand the connection between the absolute quantum phase defined in ${\cal H}_{N}^{+}$ 
(see Eq.~(\ref{p1})) and the one defined in ${\cal H}_{N}^{-}$ (see Eq.~(\ref{p2})) let us discuss briefly the absolute 
quantum angle of a plane rotator. As we have shown in the beginning, any periodic function
of the angle operator ${\bf\Theta}$ has a self-adjoint extension. In particular, the absolute 
value of the angle operator can be written in the form:
\begin{equation}
|{\bf\Theta}|=\int_{-\pi}^{\pi}d\theta\;|\theta|\;|\theta\rangle\langle\theta |
=\int_{0}^{\pi}d\theta\;\theta\;(|{\cal E}\rangle\langle {\cal E}|
+|{\cal O}\rangle\langle {\cal O}|)\;,
\label{rhs}
\end{equation}
where $|{\cal E}\rangle =(|\theta\rangle+|-\theta\rangle)/\sqrt{2}$ and
$|{\cal O}\rangle =(|\theta\rangle-|-\theta\rangle)/\sqrt{2}$.

Let us define the subspaces ${\cal H}_{e}$ and ${\cal H}_{o}$ consisting of 
all the even and odd functions of $\theta$, respectively. These Hilbert spaces are 
the analogs of ${\cal H}_{N}^{+}$ and ${\cal H}_{N}^{-}$. Therefore, the first term on the 
RHS of Eq.~(\ref{rhs}) is the analog of the absolute quantum phase defined in ${\cal H}_{N}^{+}$ 
(see Eq.~(\ref{p1})))
and the second term is the analog of the absolute quantum phase defined in ${\cal H}_{N}^{-}$ 
(see Eq.~(\ref{p2})).
On the other hand, the subspaces ${\cal H}_{e}$ and ${\cal H}_{o}$ are orthogonal because
$\langle {\cal O}|{\cal E}\rangle =0$, whereas ${\cal H}_{N}^{+}$ and ${\cal H}_{N}^{-}$ are 
not orthogonal. This is the source of the quantum phase problem, and is also the reason why we 
were able to define self-adjoint operators in ${\cal H}_{N}^{+}$ and ${\cal H}_{N}^{-}$, 
but {\it not} in ${\cal H}_{N}={\cal H}_{N}^{+}\cup {\cal H}_{N}^{-}$.

To summarize, in the case of a plane rotator, 
the z-component of the angular momentum ${\bf J}_{z}$ has a self-adjoint extension 
${\bf J}_{z}^{\alpha}$ with a domain ${\cal H}_{\alpha}$. Therefore, the condition for a 
bounded coordinate (condition~({\it i}) in the text)
implies that only periodic functions of the angle operator, ${\bf\Theta}$, are self-adjoint 
operators in ${\cal H}_{\alpha}$. In a similar manner, the lower bound for
the energy or the particle number (condition~({\it ii}) in the text) imposes that only real 
and even functions of the time or phase operator have self adjoint extensions.

\section*{Acknowledgments}
G.G. research is supported by the Killam Trust;       
F.K. research is supported by NSERC; M.R. 
research is supported by the Fund for Promotion of
Research at the Technion, and by the Technion VPR 
Fund-Glasberg-Klein Research Fund.

\end{document}